\documentclass[longauth,traditabstract]{aa} 

\usepackage{epsfig}
\usepackage{epstopdf}
\usepackage{amsmath,bm}
\usepackage{amssymb}
\usepackage[colorlinks=false,breaklinks=true]{hyperref}
\usepackage{dblfloatfix} %bottom starred figures
\usepackage{graphicx}
\usepackage{pifont}
\usepackage{stmaryrd}
\usepackage{subfigure}
\usepackage{float}
\usepackage[usenames]{color}
\usepackage{color,epsfig,amsmath,amssymb,fancyhdr}   
\usepackage{ulem}   
\usepackage{url,hyperref}
\usepackage{footmisc}
\usepackage{tabularx}
\usepackage{natbib}
\bibpunct{(}{)}{;}{a}{}{,}    %% natbib format like A&A and ApJ
\usepackage{siunitx}
\usepackage[switch]{lineno}
\newcommand{\hess}{H.E.S.S.~}
\newcommand{\g}{$\gamma$}

\begin{document}

\title{Measurement of the EBL spectral energy distribution using the VHE \g-ray spectra of H.E.S.S. blazars}

\authorrunning{\hess collaboration}
\titlerunning{Measurement of the EBL SED with \hess}

\author{\small{H.E.S.S. Collaboration
\and H.~Abdalla \inst{1}
\and A.~Abramowski \inst{2}
\and F.~Aharonian \inst{3,4,5}
\and F.~Ait~Benkhali \inst{3}
\and A.G.~Akhperjanian\protect\footnotemark[2] \inst{6,5} % Deceased author
\and T.~Andersson \inst{10}
\and E.O.~Ang\"uner \inst{21}
\and M.~Arakawa \inst{43}
\and M.~Arrieta \inst{15}
% ...
}}

\institute{
Centre for Space Research, North-West University, Potchefstroom 2520, South Africa \and 
Universit\"at Hamburg, Institut f\"ur Experimentalphysik, Luruper Chaussee 149, D 22761 Hamburg, Germany \and 
Max-Planck-Institut f\"ur Kernphysik, P.O. Box 103980, D 69029 Heidelberg, Germany \and 
Dublin Institute for Advanced Studies, 31 Fitzwilliam Place, Dublin 2, Ireland \and 
% 5
National Academy of Sciences of the Republic of Armenia,  Marshall Baghramian Avenue, 24, 0019 Yerevan, Republic of Armenia  \and
Yerevan Physics Institute, 2 Alikhanian Brothers St., 375036 Yerevan, Armenia \and
Institut f\"ur Physik, Humboldt-Universit\"at zu Berlin, Newtonstr. 15, D 12489 Berlin, Germany \and
University of Namibia, Department of Physics, Private Bag 13301, Windhoek, Namibia \and
GRAPPA, Anton Pannekoek Institute for Astronomy, University of Amsterdam,  Science Park 904, 1098 XH Amsterdam, The Netherlands \and
% 10
Department of Physics and Electrical Engineering, Linnaeus University,  351 95 V\"axj\"o, Sweden \and
Institut f\"ur Theoretische Physik, Lehrstuhl IV: Weltraum und Astrophysik, Ruhr-Universit\"at Bochum, D 44780 Bochum, Germany \and
GRAPPA, Anton Pannekoek Institute for Astronomy and Institute of High-Energy Physics, University of Amsterdam,  Science Park 904, 1098 XH Amsterdam, The Netherlands \and
Institut f\"ur Astro- und Teilchenphysik, Leopold-Franzens-Universit\"at Innsbruck, A-6020 Innsbruck, Austria \and
School of Physical Sciences, University of Adelaide, Adelaide 5005, Australia \and
% 15
LUTH, Observatoire de Paris, PSL Research University, CNRS, Universit\'e Paris Diderot, 5 Place Jules Janssen, 92190 Meudon, France \and
Sorbonne Universit\'es, UPMC Universit\'e Paris 06, Universit\'e Paris Diderot, Sorbonne Paris Cit\'e, CNRS, Laboratoire de Physique Nucl\'eaire et de Hautes Energies (LPNHE), 4 place Jussieu, F-75252, Paris Cedex 5, France \and
Laboratoire Univers et Particules de Montpellier, Universit\'e Montpellier, CNRS/IN2P3,  CC 72, Place Eug\`ene Bataillon, F-34095 Montpellier Cedex 5, France \and
DSM/Irfu, CEA Saclay, F-91191 Gif-Sur-Yvette Cedex, France \and
Astronomical Observatory, The University of Warsaw, Al. Ujazdowskie 4, 00-478 Warsaw, Poland \and
% 20
Aix Marseille Universit\'e, CNRS/IN2P3, CPPM UMR 7346,  13288 Marseille, France \and
Instytut Fizyki J\c{a}drowej PAN, ul. Radzikowskiego 152, 31-342 Krak{\'o}w, Poland \and
Funded by EU FP7 Marie Curie, grant agreement No. PIEF-GA-2012-332350,  \and
School of Physics, University of the Witwatersrand, 1 Jan Smuts Avenue, Braamfontein, Johannesburg, 2050 South Africa \and
Laboratoire d'Annecy-le-Vieux de Physique des Particules, Universit\'{e} Savoie Mont-Blanc, CNRS/IN2P3, F-74941 Annecy-le-Vieux, France \and
% 25
Landessternwarte, Universit\"at Heidelberg, K\"onigstuhl, D 69117 Heidelberg, Germany \and
Universit\'e Bordeaux, CNRS/IN2P3, Centre d'\'Etudes Nucl\'eaires de Bordeaux Gradignan, 33175 Gradignan, France \and
Oskar Klein Centre, Department of Physics, Stockholm University, Albanova University Center, SE-10691 Stockholm, Sweden \and
Wallenberg Academy Fellow,  \and
Institut f\"ur Astronomie und Astrophysik, Universit\"at T\"ubingen, Sand 1, D 72076 T\"ubingen, Germany \and
% 30
Laboratoire Leprince-Ringuet, Ecole Polytechnique, CNRS/IN2P3, F-91128 Palaiseau, France \and
APC, AstroParticule et Cosmologie, Universit\'{e} Paris Diderot, CNRS/IN2P3, CEA/Irfu, Observatoire de Paris, Sorbonne Paris Cit\'{e}, 10, rue Alice Domon et L\'{e}onie Duquet, 75205 Paris Cedex 13, France \and
Univ. Grenoble Alpes, IPAG,  F-38000 Grenoble, France \protect\\ CNRS, IPAG, F-38000 Grenoble, France \and
Department of Physics and Astronomy, The University of Leicester, University Road, Leicester, LE1 7RH, United Kingdom \and
Nicolaus Copernicus Astronomical Center, Polish Academy of Sciences, ul. Bartycka 18, 00-716 Warsaw, Poland \and
% 35
Institut f\"ur Physik und Astronomie, Universit\"at Potsdam,  Karl-Liebknecht-Strasse 24/25, D 14476 Potsdam, Germany \and
Friedrich-Alexander-Universit\"at Erlangen-N\"urnberg, Erlangen Centre for Astroparticle Physics, Erwin-Rommel-Str. 1, D 91058 Erlangen, Germany \and
DESY, D-15738 Zeuthen, Germany \and
Obserwatorium Astronomiczne, Uniwersytet Jagiello{\'n}ski, ul. Orla 171, 30-244 Krak{\'o}w, Poland \and
Centre for Astronomy, Faculty of Physics, Astronomy and Informatics, Nicolaus Copernicus University,  Grudziadzka 5, 87-100 Torun, Poland \and
% 40
Department of Physics, University of the Free State,  PO Box 339, Bloemfontein 9300, South Africa \and
Heisenberg Fellow (DFG), ITA Universit\"at Heidelberg, Germany  \and
GRAPPA, Institute of High-Energy Physics, University of Amsterdam,  Science Park 904, 1098 XH Amsterdam, The Netherlands \and
Department of Physics, Rikkyo University, 3-34-1 Nishi-Ikebukuro, Toshima-ku, Tokyo 171-8501, Japan
% ...
}

\offprints{H.E.S.S.~collaboration,
\protect\\\email{\href{mailto:contact.hess@hess-experiment.eu}{contact.hess@hess-experiment.eu}};
\protect\\\protect\footnotemark[1] Corresponding authors
\protect\\\protect\footnotemark[2] Deceased
}

\abstract{
Very high-energy \g-rays (VHE, $E \gtrsim 100$ GeV) propagating over cosmological distances can interact with the low-energy photons of the extragalactic background light (EBL) and produce electron-positron pairs. The transparency of the universe to VHE \g-rays is then directly related to the spectral energy distribution (SED) of the EBL. The observation of features in the VHE energy spectra of extragalactic sources allows the EBL to be measured, which otherwise is very difficult to determine. An EBL-model independent measurement of the EBL SED with the \hess array of Cherenkov telescopes is presented. It is obtained by extracting the EBL absorption signal from the reanalysis of high-quality spectra of blazars. From \hess data alone the EBL signature is detected at a significance of 9.5$\sigma$, and the intensity of the EBL obtained in different spectral bands is presented together with the associated \g-ray horizon.
}

\keywords{Gamma rays: galaxies ---  Cosmology: cosmic background radiation} 

\maketitle
%\makeatletter
%\renewcommand*{\@fnsymbol}[1]{\ifcase#1\@arabic{#1}\fi}
%\makeatother

%\onecolumn
%\tableofcontents

%\twocolumn
%\newpage

\section{Introduction}
The extragalactic background light (EBL) is the second most intense background photon field in the universe after the cosmic microwave background. This diffuse radiation stems from the integrated light emitted through thermal and non-thermal processes, and its reprocessing by the interstellar medium over cosmic history. It covers wavelengths ranging from the ultraviolet to far infrared and submillimeter wavelengths. Direct measurements of the EBL are very difficult because of foreground contamination due to zodiacal light and diffuse Galactic light \citep{Hauser:1998ri}. Lower limits have been derived from galaxy counts, and models have been developed to describe its spectral energy distribution (SED) (see \textit{e.g.} \citealp{Franceschini:2008tp, Dominguez:2010bv, 2010A&A...515A..19K, Finke:2009xi, Gilmore:2011ks}). This SED is usually described with two main components: an optical component due to starlight emission and an infrared component due to the reprocessing of starlight by dust. The EBL SED contains unique information about galaxy formation and evolution. Its study is therefore of interest for cosmology. This low-energy photon background is responsible for the limited horizon of very high-energy (VHE, $E \gtrsim 100$ GeV) photons, since these \g-rays interact with EBL photons through the production of electron-positron pairs, resulting in attenuated observed fluxes above the reaction threshold \citep{Nikishov62, Gould:1967zzb}. While this affects the study of extragalactic \g-ray sources, it also provides a way to probe the EBL itself (for reviews see \textit{e.g.} \citealp{Hauser:2001xs, Dwek:2012nb, Costamante:2013sva}).\\

The attenuation of \g-rays on the EBL is an energy-dependent process which leads to a specific spectral signature. Observations of features in the VHE spectra of extragalactic sources with Cherenkov telescopes like H.E.S.S. can thus be used to constrain the EBL under some assumptions on the intrinsic spectra of the considered sources. Indeed, a major complication in constraining the EBL with \g-rays comes from the indeterminacy of the intrinsic energy spectra of sources, and consequently there is a possible degeneracy between intrinsic curvature and EBL attenuation.
The technique that was first applied to constrain the EBL with \hess relied on the assumption of a theoretical limit for the hardness of the intrinsic power-law spectra. Upper limits on the level of EBL were obtained given the softness of the observed spectra. Using the two blazars H 2356$-$309 ($z=0.165$) and 1ES~1101$-$232 ($z=0.186$), \hess  showed that the universe was more transparent to \g-rays than expected at that time from direct EBL measurements \citep{Aharonian:2005gh, Matsumoto:2004dx}. The upper limits on the EBL density thereby obtained with \hess turned out to be close to the lower limits derived from galaxy counts. A global reassessment of EBL models followed. This \hess study was followed by a model-dependent determination of the EBL \citep{Abramowski:2012ry} obtained by simultaneously fitting the EBL optical depths and intrinsic spectra of a sample of extragalactic sources with a maximum-likelihood method assuming smooth concave intrinsic shapes. The shape of the EBL SED was frozen choosing the model given in \citet{Franceschini:2008tp}, leaving only the normalization free. The overall test statistic led to an 8.8$\sigma$ detection of EBL absorption with respect to no absorption, with a normalization factor relative to this model of $1.27^{+0.18\ }_{-0.15\ } ({\rm stat}) \pm 0.25 ({\rm sys})$ in the 1.2 $\mu$m to 5.5 $\mu$m wavelength range. In the following, this study will be referred to as HESS2013. Other EBL constraints and measurements using \g-rays have been conducted as well \textit{e.g.} with \textit{Fermi} LAT, MAGIC and VERITAS \citep{Ackermann:2012sza, Abeysekara:2015pjl, Ahnen:2016gog}.\\

The new analysis presented in the present paper follows a different approach, focusing on the determination of the shape of the EBL SED in addition to its overall normalization. An extended sample of blazars is used with respect to HESS2013, simultaneously fitting their VHE intrinsic spectra together with a generic attenuation. As the EBL is expected to leave a typical energy-dependent and redshift-dependent imprint on the observed spectra, the detection of such a modulation can be used to translate the absorption pattern into spectrally-resolved EBL intensity levels. This analysis considers high-quality VHE spectra and assumes featureless intrinsic spectra, allowing for intrinsic curvature.
This approach aims not only for a \hess measurement of the EBL SED independently of any EBL model but also for a generic characterization of the universe's transparency to VHE \g-rays with the fewest possible priors. Beyond the interest for the EBL \textit{per se}, this is particularly relevant for the study of potential second-order processes in the propagation of \g-rays over cosmological distances. Those include for instance conversion into axion-like particles (\textit{e.g.} \citealp{SanchezConde:2009wu, Abramowski:2013oea}), Lorentz invariance violation (\textit{e.g.} \citealp{Stecker:2001vb, Jacob:2008gj}) or cascade emission in extragalactic magnetic fields (\textit{e.g.} \citealp{1994ApJ...423L...5A, Taylor:2011bn}).\\

Spectral features are searched for in the reanalysis of \hess phase-I (four-telescope) data with a new method to measure the EBL SED. Using only H.E.S.S. data offers the possibility to handle systematic uncertainties from different spectra in a homogeneous and well-controlled way. Furthermore, published spectral points are not usually released together with their covariance matrix. 
Using this additional information, these results are expected to be more robust than similar studies using published spectra from different Cherenkov telescopes only (\textit{e.g.} \citealp{Orr2011, Meyer:2012us, Sinha:2014lfa, Biteau:2015xpa}).\\

This paper is organized as follows: The blazar sample and the data analysis are described in Sec.~\ref{Section:Hess}. In Sec.~\ref{NullHypo}, the need for an energy- and redshift-dependent modulation in the energy spectra is demonstrated. In Sec.~\ref{EBLOpticalDepth}, the EBL absorption process is presented in detail, and in Sec.~\ref{Section:Method}, the method used to translate the modulation seen in spectra in terms of EBL is described. The results are presented and discussed in Sec.~\ref{results}.

\section{H.E.S.S. data analysis}
\label{Section:Hess}

\subsection{Data reduction}
H.E.S.S. is an array of five imaging atmospheric Cherenkov telescopes located in the Khomas Highland, Namibia ($23^\circ 16'18''$ S, $16^\circ 30'01''$ E) at an elevation of 1800 m above sea level. In this work, only data from the four telescopes of the first phase of H.E.S.S. are used. This initial four-telescope array detects \g-rays above $\sim100$ GeV with an energy resolution better than $15 \%$ \citep{Aharonian:2006pe}.
Data reduction is performed using the \textit{Model Analysis} technique~\citep{deNaurois:2009ud} in which recorded air-shower images are compared to template images pre-calculated using a semi-analytic model and a log-likelihood optimization technique. For a wider energy coverage, the \textit{Loose cuts} of the \textit{Model Analysis} are adopted, corresponding to a selection criterion on the image charge of a minimum of 40 photo-electrons. A cross-check analysis, performed with the \textit{ImPACT} analysis~\citep{Parsons:2014voa} and an independent calibration chain, yields compatible results. 
 
 \subsection{Blazar sample}
The data sets used are those of blazars with known redshift observed by \hess with a high significance, listed in Table~\ref{Datasets}. They all belong to the class of high-frequency-peaked BL Lac objects. Their VHE emission is therefore not expected to be affected by the local blazar environment. Blazars can sometimes show signs of spectral variability correlated with their flux level, and this could bias the interpretation in terms of the EBL. To avoid this, data from sources with known variability are divided into subsets within logarithmic flux bins, like in HESS2013. These subsets, labeled by a number, are ordered by increasing flux level. The bulk of the data sample is similar to the one used in HESS2013, with some differences mentioned below.

Mrk 421  ($z$ = 0.031, \citealp{1975ApJ...198..261U}) is the first extragalactic source detected in the VHE domain \citep{Punch:1992xw}. This bright and variable northern-sky blazar is observed by \hess at large zenith angles ($\gtrsim 60^{\circ}$) \citep{Aharonian:2005ib}. As a consequence, the energy threshold is high ($\gtrsim 1$ TeV) due to the strong atmospheric absorption of Cherenkov light. On the other hand, the effective area is relatively large at higher energies, resulting in a spectrum extending above 10 TeV. In addition to the 2004 observations on Mrk 421 (labels 1 to 3), data taken during the 2010 high state \citep{Tluczykont:2011gs} are added (labels 4 and 5). Mrk 501 ($z$ = 0.034, \citealp{moles1987two}) is the second extragalactic VHE source detected \citep{Quinn:1996dj} and is also observed by \hess at large zenith angles. Data taken during the 2014 high state are used \citep{Cologna:2015mia}. These low-redshift spectra at multi-TeV energies are key to probe the mid-IR region of the EBL spectrum.
For PKS~2005$-$489 ($z$ = 0.071, \citealp{1987ApJ...318L..39F}), the data used here are identical to that used in HESS2013, and detailed in \citet{PKS2005_HESS2010} and \citet{PKS2005_HESS2011}.
PKS2155$-$304 ($z$ = 0.116, \citealp{1993ApJ...411L..63F}) is a very bright blazar extensively studied by \hess As in HESS2013, the data of the exceptional July 2006 high state are used (labels 1 to 7), together with observations of the 2008 low state (label 8). These very-high-significance data sets yield excellent quality spectra that are crucial for an unambiguous identification of the EBL-absorption pattern.
For 1ES 0229$+$200 ($z$ = 0.14, \citealp{1993ApJ...412..541S}), H 2356$-$309 ($z$ = 0.165, \citealp{Jones:2009yz}), 1ES~1101$-$232 ($z$ = 0.186, \citealp{1989ApJ...345..140R}) and 1ES~0347$-$121 ($z$ = 0.188, \citealp{Woo:2005hy}), the data are identical to that used in HESS2013.
1ES 0414$+$009 ($z$ = 0.287, \citealp{1991AJ....101..818H}) is also added to the sample. This distant blazar was observed by \hess from 2005 to 2009 \citep{HESS20121ES0414}.

\begin{table*}
\caption{Properties of the data sets used in this study, including the observation live time, the detection significance~$\sigma$, the source redshift $z$, and the energy range covered by the unfolded $\gamma$-ray spectra.}
\begin{center}
%\begin{tabular}{S[table-format=2.0]SSSS[table-format = 0.2]}
\begin{tabular}{l S  @\qquad S @\qquad S @\qquad c}
 Data set & $\text{Live time}$  & $\sigma$ \ \ \ & $z$  \ & $E_{\rm min} - E_{\rm max}$ \\
 & $\text{(hours)}$ & & &  (TeV) \\ 
  \hline
  Mrk 421 (1) & 4.9 & 89.6 & 0.031 & $1.41 -14.9$ \\
  Mrk 421 (2) & 3.8 & 122 & 0.031 & $1.22 -15.9$ \\
  Mrk 421 (3) & 2.9 & 123 & 0.031 & $1.19 -19.5$ \\
  Mrk 421 (4) & 3.3 & 96.2 & 0.031 & $1.6 -16.5$ \\
  Mrk 421 (5) & 1.6 & 46.0 & 0.031 & $1.5 -15.2$ \\
  Mrk 501  & 1.8 & 66.7 & 0.034 & $1.9 -19.5$ \\
  PKS 2005$-$489 (1) & 71.2 & 28.8 & 0.071 & $0.29 - 1.6$\\
  PKS 2005$-$489 (2) & 18.7 & 29.2 & 0.071 & $0.29 - 3.0$\\
  PKS 2155$-$304 (1) & 7.4 & 94.8 & 0.116 & $0.24 - 4.6$\\
  PKS 2155$-$304 (2) & 6.1 & 119 & 0.116 & $0.24 - 1.98$\\
  PKS 2155$-$304 (3) & 5.5 & 187 & 0.116 & $0.24 - 3.7$\\
  PKS 2155$-$304 (4) & 2.6 & 135 & 0.116 & $0.24 - 2.44$\\
  PKS 2155$-$304 (5) & 3.5 & 227 & 0.116 & $0.24 - 4.6$\\
  PKS 2155$-$304 (6) & 1.3 & 172 & 0.116 & $0.29 - 4.6$\\
  PKS 2155$-$304 (7) & 1.3 & 200 & 0.116 & $0.29 - 3.6$\\
  PKS 2155$-$304 (8) & 25.4 & 111 & 0.116 & $0.19 - 3.7$\\
  1ES 0229$+$200 & 57.7 & 11.6 & 0.14 & $0.4 - 2.8$\\
  H 2356$-$309 & 92.6 & 19.6 & 0.165 & $0.19 - 1.98$\\
  1ES 1101$-$232 & 58.2 & 16.8 & 0.186 & $0.19 - 1.98$ \\
  1ES 0347$-$121 & 33.9 & 14.1 & 0.188 & $0.19 - 6.9$ \\
  1ES 0414$+$009 & 73.7 & 9.6 & 0.287 & $0.19 - 0.69$ \\
   \hline
\end{tabular}
\end{center}
\label{Datasets}
\end{table*}

\subsection{Spectral deconvolution}
Published spectra by \hess are usually obtained by means of a forward-folding method \citep{Piron:2001ir} for which an assumption on the spectral shape is required. The results of such a procedure are spectral parameters and their associated errors. Spectral points can then be constructed in different energy bins (using the ratio of the observed signal to the signal predicted by the fitted shape in each bin), but such points are not a direct result of the forward-folding procedure. The present study follows a different approach: the energy spectrum of each data set is obtained using a Bayesian unfolding technique based on \citet{Albert:2007qw} (and already used by \hess in \citealt{2013AxionPaper}) in order to directly obtain spectral points independently of any \textit{a priori} spectral shape, together with their correlations. This is a key aspect of this new analysis since these unfolded spectra allow the exploration of spectral patterns.

The energy threshold used in the spectral deconvolution is defined as the energy at which the effective area reaches $15\%$ of its maximum value. This is a standard procedure used in \hess \citep{2014AGNUL, HESS2FAGN}.
A fixed logarithmic binning in energy is chosen for the deconvolution of each spectrum, adapted to the energy resolution, and a minimum significance of $2 \sigma$ per bin is required for a spectral point to be defined. The high-energy end of the range indicated in Table~\ref{Datasets} reflects the tail of the significance distribution in energy. The consistency of the unfolded spectral points have been verified to be in excellent agreement with the residual points of the above-mentioned forward folding procedure.
The 21 unfolded spectra of the sample yield a total of 247 spectral points.

\section{The null hypothesis: fits without EBL}
\label{NullHypo}
This determination of the EBL is based on the observation of features in observed VHE spectra. Assumptions made on intrinsic spectra are therefore essential. In this section, these assumptions are presented, and it is shown that the assumption of featureless spectra with no EBL leads to a poor fit of the data. This calls for additional degrees of freedom in the data interpretation. The best fit without EBL will be later considered as the null hypothesis. In the next sections, it will be shown that when these additional degrees of freedom reflect EBL levels in different bands, the fit is significantly improved and interpreted as evidence for a spectrally-resolved EBL detection.

The simplest description of the energy spectrum of a non-thermal \g-ray source like a blazar is the two-parameter power-law function 

\begin{equation}
\Phi_{\rm{PWL}}(E_\gamma) = \Phi_0 (E_\gamma/ E_0)^{-\alpha}.
\end{equation}

\begin{figure*}[]
\centering
\subfigure[\label{SubFig_1}]{\includegraphics[scale=0.29]{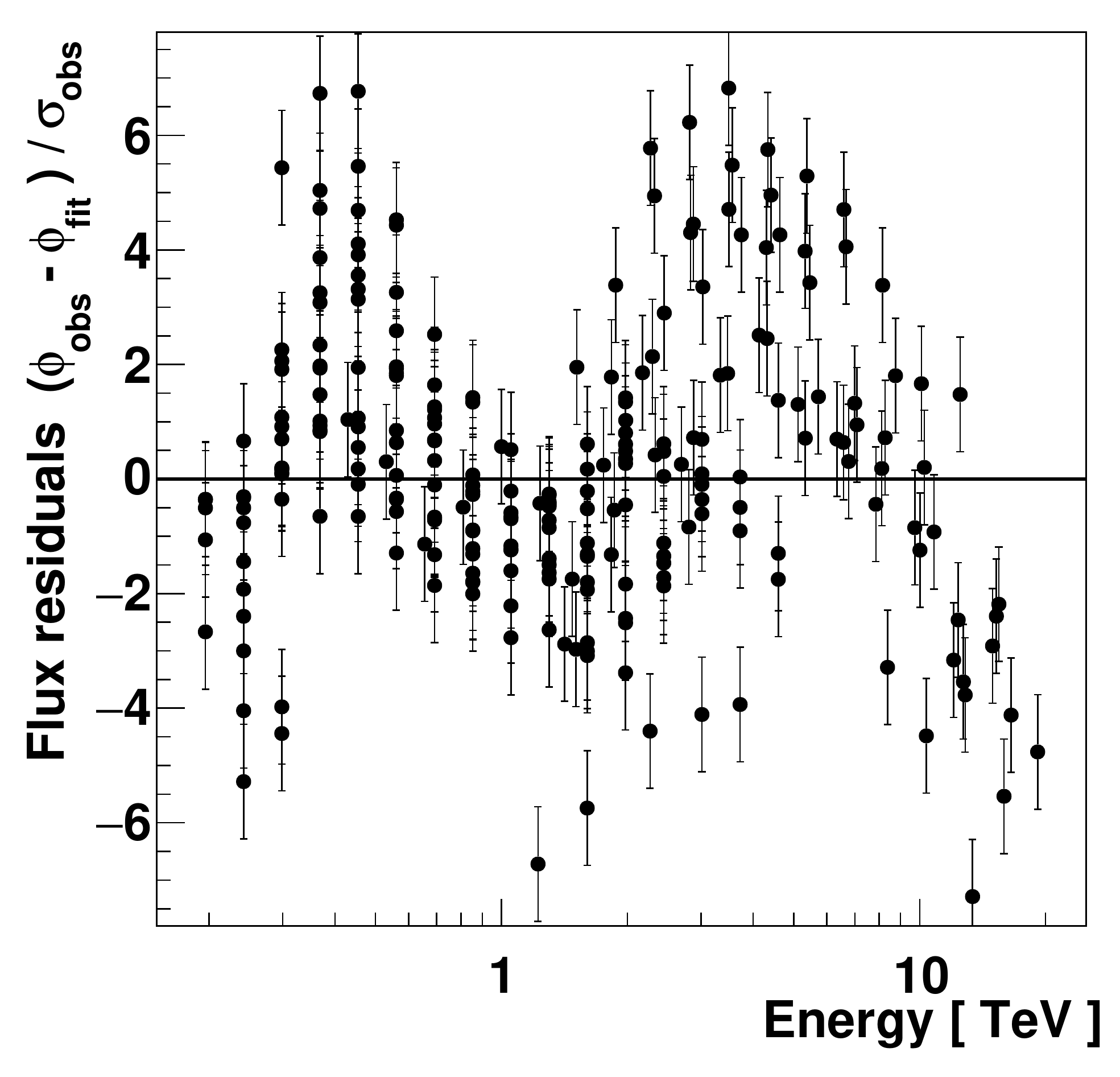}}
\subfigure[\label{SubFig_2}]{\includegraphics[scale=0.29]{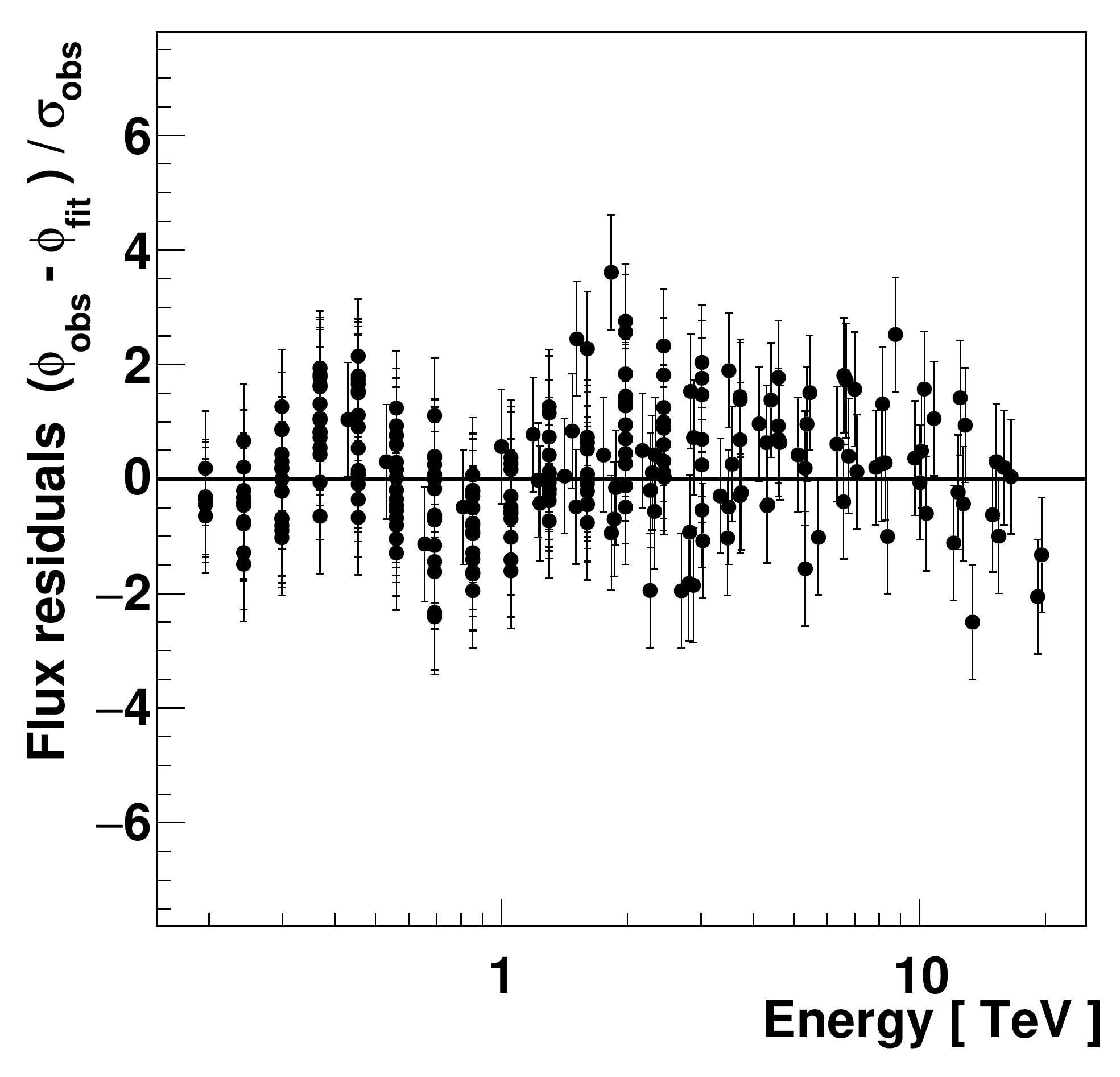}}
\subfigure[\label{SubFig_3}]{\includegraphics[scale=0.29]{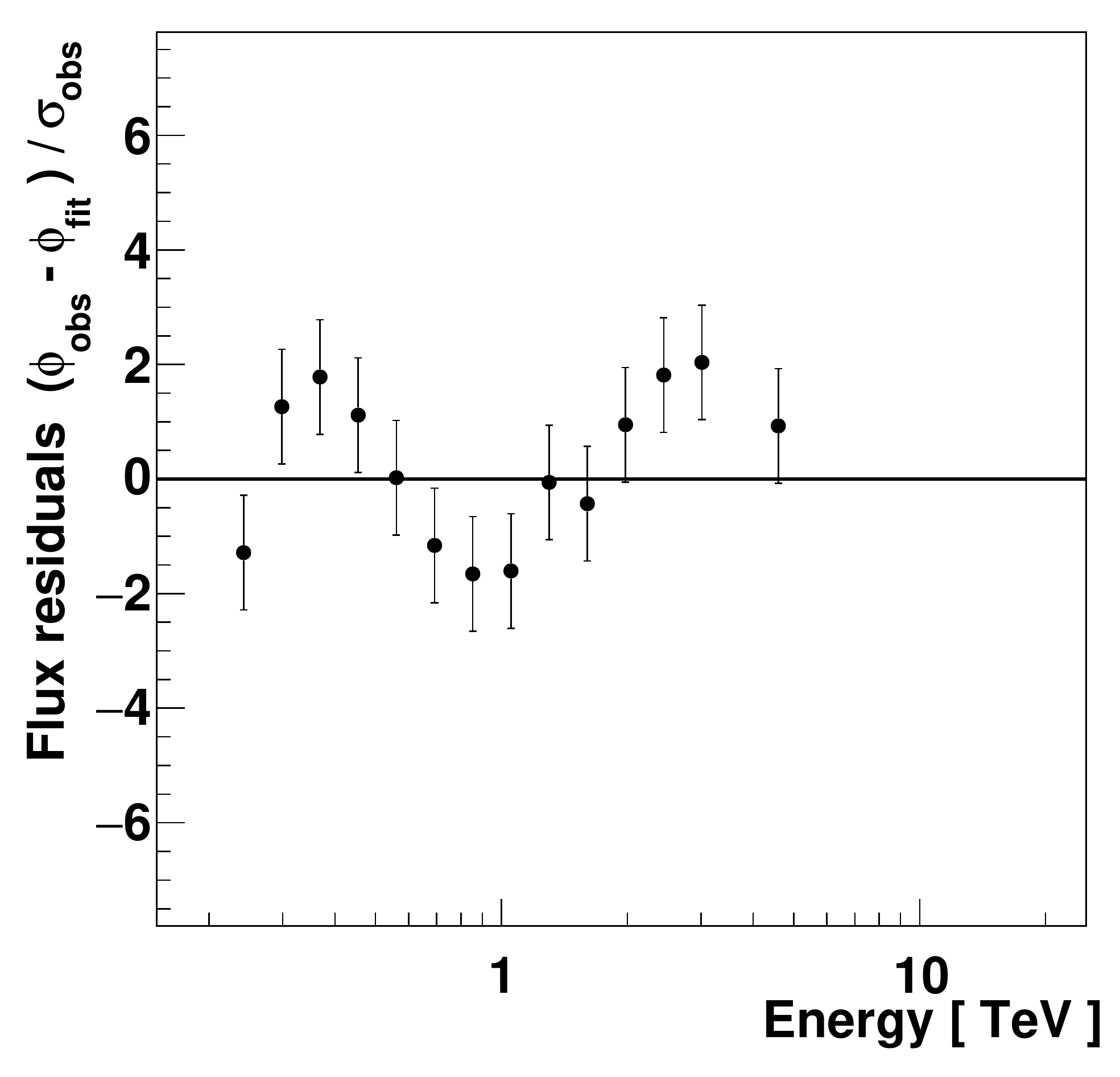}}
\label{NoEBLResiduals}
\caption{Fit residuals of the featureless spectral shapes, as a function of energy. \subref{SubFig_1}~Residuals of the whole sample of spectra to the power-law fit. \subref{SubFig_2}~Residuals of the whole sample to the log-parabola fit. \subref{SubFig_3}~Example residuals to the log-parabola fit for the subset PKS~2155$-$304~(5).}
\end{figure*}

 Fitting each spectrum $j$ of the blazar sample with a power-law yields overall $\sum_j \chi^2_{j,\rm{ PWL}}=1472.8$, for $205$ degrees of freedom. The fit residuals have a large dispersion, as shown in Fig.~\ref{SubFig_1}. In addition, a visible modulation indicates the need for a more elaborate parameterization. Spectral curvature is introduced with the three-parameter log-parabola function 
 
\begin{equation}
\Phi_{\rm{LP}}(E_\gamma) = \Phi_0 (E_\gamma/ E_0)^{-\alpha - \beta \log(E_\gamma/ E_0)},
\end{equation}

  with $\beta \ge 0$ the additional curvature parameter. The use of the log-parabola function improves the fit significantly (at the 30$\sigma$ level) as $\sum_j \chi^2_{j,\rm{ LP}}=281.07$, for $184$ degrees of freedom. However, a modulation in the distribution of the fit residuals is still present, as illustrated in Fig.~\ref{SubFig_2}. As an example, Fig.~\ref{SubFig_3} shows a residual distribution isolated from Fig.~\ref{SubFig_2}, for the log-parabola fit of the subset PKS~2155-304 (5).

The same observation can be made using other featureless spectral shapes, as long as no intrinsic irregularities are considered. It is then natural to aim at interpreting these modulations of the flux residuals as the effect of EBL attenuation. The properties of these energy-dependent modulations and their redshift dependence are the central point of this study to measure the EBL SED.

In the following, these modulations -- not accounted for by featureless intrinsic shapes --  are translated in terms of spectrally-resolved EBL levels.

\section{EBL optical depth}
\label{EBLOpticalDepth}
The extragalactic medium at a given redshift $z$ is filled with EBL photons of proper number density $n_{\mbox{\tiny{EBL}}}(\epsilon,z)$ at proper energy $\epsilon$. The opacity of this medium for \g-rays of observed energy $E_\gamma$ coming from a source at redshift $z_s$ is encoded in the optical depth $\tau(E_\gamma,z_s)$  \citep{Gould:1967zzb, Stecker:1992wi}. It consists of an integration over $z$, $\epsilon$ and the angle $\theta$ between the photon momenta:

\begin{align}
\tau(E_\gamma,z_s) =&  \int_0^{z_s} dz \frac{dl}{dz} \int_{\epsilon_{thr}}^\infty d \epsilon \frac{dn_{\mbox{\tiny{EBL}}}}{d\epsilon}(\epsilon,z) \nonumber \\ & \int_0^2 d\mu \frac{\mu}{2} \ \sigma_{\gamma \gamma} \left[\beta(E_\gamma(1+z), \epsilon, \mu )\right],
\label{tau}
\end{align}
where $\mu = 1-\cos(\theta) $, and $\epsilon_{thr} (E_\gamma,z)= \frac{2 m_e^2 c^4}{ E_\gamma \mu(1+z)}$ is the threshold energy dictated by kinematics. The cross section for pair production \citep{Breit:1934zz, Gould:1967zzb} is defined as:

\begin{equation}
\sigma_{\gamma \gamma}(\beta)=\frac{3 \sigma_T}{16}(1-\beta^2)\left[(3-\beta^4) \ln(\frac{1+\beta}{1-\beta}) - 2 \beta (2-\beta^2)\right],
\end{equation}
where  $\beta c $ is the velocity of the outgoing electron and positron in the center of mass system, and $\sigma_T$ is the Thomson cross section. \\

A flat $\Lambda$CDM cosmology with Hubble constant $H_0$, matter density parameter $\Omega_M$ and dark energy density parameter $\Omega_{\Lambda}$ is considered. The distance element in Eq.~\ref{tau} then reads

\begin{equation}
\frac{dl}{dz}= c \left( H_0 (1+z) \sqrt{\Omega_M(1+z)^3 +\Omega_\Lambda}\right) ^{-1},
\end{equation}
where the values $H_0 = 70 \text{ km}  \text{ s}^{-1}  \text{Mpc}^{-1}$, $\Omega_M= 0.3 $ and $\Omega_{\Lambda} = 0.7$ are assumed. The most influential cosmological parameter is the Hubble constant as $\tau$ scales linearly with $1/H_0$. This generic choice of $H_0$ is in line with the latest Planck results obtained from cosmic microwave background data \citep{Ade:2015xua} and with the results obtained from more local constraints \citep{Riess:2016jrr}. 
The dependence of the results on the precise choice for $H_0$ is negligible with respect to the sensitivity of the method. For a detailed study on the influence of cosmological parameters on \g-ray attenuation see \textit{e.g.} \citet{Dominguez:2013mfa}.

The evolution of the EBL in Eq. \ref{tau} with redshift is accounted for decoupling the local ($z=0$) EBL SED and an evolution function, 

\begin{equation}
d \epsilon \frac{dn_{\mbox{\tiny{EBL}}}}{d\epsilon}(\epsilon,z) =  d \epsilon_0 \frac{dn_{\mbox{\tiny{EBL}}}}{d\epsilon_0}(\epsilon_0,0) \times f(\epsilon_0, z),
\label{EBlevol}
\end{equation}
where $\epsilon_0 = \epsilon/(1+z)$ is the EBL energy at $z=0$. The evolution function $f(\epsilon_0, z)$ is extracted from the model given in \citet{Franceschini:2008tp} using the ratio of the SED at a redshift $z$ to its value at $z=0$. The influence on the results of this model-dependent ingredient for the EBL evolution with redshift is weak, as discussed in Sec.~\ref{Section:Sys}.

The observed energy spectrum $\Phi_{\mathrm{obs}}(E_\gamma)$ of an extragalactic source is the convolution of its intrinsic spectrum $\Phi_{\mathrm{int}}(E_\gamma)$ with the EBL absorption effect $\Phi_{\mathrm{obs}}(E_\gamma)= \Phi_{\mathrm{int}}(E_\gamma) \times e^{-\tau(E_\gamma,z_s)}$. EBL absorption then leaves a redshift and energy-dependent imprint on the observed VHE spectra of blazars.

\section{Method}
\label{Section:Method}
 
\subsection{Parameterization of the EBL SED and intrinsic blazar spectra}
As shown in Sec.~\ref{NullHypo}, energy-dependent modulations in the residuals of spectral fits with featureless functions call for additional degrees of freedom. Those can be interpreted in terms of EBL absorption. A determination of the EBL is possible by confronting \g-ray data with different EBL hypotheses, and in the present approach these hypotheses are required to be independent of EBL models. 
A preliminary study testing EBL shapes as splines constructed upon a grid in energy density (like in \citealt{Mazin:2007pn}) showed indeed that the shape of the EBL was accessible with H.E.S.S. data \citep{Lorentz:2015msa}.
In the present study a different and more robust method is used: EBL-related degrees of freedom are introduced as continuous levels of EBL intensity in different bands over the range of interest for \g-ray absorption. This approach allows a more accurate estimation of uncertainties and a more meaningful statistical treatment of data sets as compared to the use of splines on a grid.

The local EBL energy density is decomposed into connected energy bands with bounds $\left[ \epsilon_i, \epsilon_{i+1} \right]$ and content $\rho_i \ge 0$ 

\begin{align}
& \epsilon_0^2 \frac{dn_{\mbox{\tiny{EBL}}}}{d\epsilon_0} =  \sum_i w_i(\epsilon_0) \rho_i, \nonumber \\ & \text{where } w_i(\epsilon_0) = \left\{
    \begin{array}{ll}
        1 \text{ if } \epsilon_0 \in \left[ \epsilon_i, \epsilon_{i+1} \right] \\
        0 \text{ otherwise} 
    \end{array}
\right ..
\label{EBLPerBin}
\end{align}
 
This parameterization of the EBL SED is injected into the optical depth calculation (Eq. \ref{tau}). The set of EBL levels $\left\lbrace \rho_i \right\rbrace $ is adjusted to fit the absorption pattern in \g-ray data. This approach is similar to the one used in \citet{1998PhRvL..80.2992B} to derive upper limits on the EBL SED. 

The local EBL SED  is divided into four bands in energy (wavelength) with equal-size logarithmic widths. This simple choice is found to be optimal in terms of sensitivity. Increasing the number of subdivisions does not significantly improve the fit quality but leads to an increase of the errors on the EBL levels.
The low-energy bound  $\epsilon_{0, \rm{min}}$ (or equivalently the high-wavelength bound $\lambda_{0, \rm{max}}$) of the local EBL range corresponds to the threshold for pair creation with the most energetic \g-rays of the blazar sample following the threshold relation previously mentioned:

\begin{equation}
\epsilon_{0, \rm{min}} =  \frac{\hbar c}{\lambda_{0, \rm{max}}} =  \frac{2 m_e^2 c^4}{E_\gamma \mu (1+z_s)^2}.
\label{MaxThreshold}
\end{equation}

The high-energy (low-wavelength) bound of the EBL range is chosen beyond the peak of the cross section for interaction with \g-rays in the lowest energy spectral point of the sample and adjusted \textit{a posteriori} as the energy at which the sensitivity in this band is seen to decrease.\\

Intrinsic spectral shapes are described with log-parabolas. This naturally includes power laws in cases where the fit prefers vanishing curvature parameters.
This choice is intended to avoid attributing the entire origin of spectral curvature to the EBL. As only positive values of $\alpha$ and $\beta$ are considered, another implicit assumption is the non-convexity of the intrinsic spectra, like in \citet{2005ApJ...618..657D,2005ApJ...634..155D}. These simple considerations ensure that this EBL determination does not rely on specific assumptions about the underlying acceleration mechanism behind the VHE \g-ray emission of blazars. The consideration of more complex intrinsic shapes does not lead to an improvement of individual fit qualities. 

\subsection{Joint fit}

For each individual data set, a joint fit of the EBL levels and intrinsic spectral parameters is performed. The covariance matrix $C_\mathrm{T}$ determined in the unfolding procedure is used in order to take into account the correlations between spectral points in the $\chi^2$ minimization. The minimized function is 

\begin{equation}
\chi^2 = (\vec{\Phi}_{\text{test}} - \vec{\Phi}_{\text{obs}} )^T C_\mathrm{T}^{-1} (\vec{\Phi}_{\text{test}} - \vec{\Phi}_{\text{obs}} ) ,
\label{Chi2Expression}
\end{equation}
where $\vec{\Phi}_{\mathrm{obs}}$ is the vector of observed spectral points and $\vec{\Phi}_{\mathrm{test}}$ is the vector of tested functions. Those test functions include both EBL parameters $\left\lbrace \rho_i \right\rbrace$ and intrinsic spectral parameters ($\Phi_0, \alpha, \beta$):

\begin{equation}
\Phi_{\mathrm{test}}(E_\gamma, z_s)= \Phi_{\mathrm{int}}(E_\gamma,\Phi_0, \alpha, \beta ) \times e^{-\tau(E_\gamma,z_s,\left\lbrace \rho_i \right\rbrace )},
\label{FitFunction}
\end{equation}
resulting in a seven-parameter fit. 

The four EBL levels are independent in the fit of each individual spectrum and are combined later, as described in Sec. \ref{CombineSection}. This approach allows a clear identification of the contribution of each spectrum to the overall results, at different wavelengths.

\section{Results}
\label{results}
Fits of all spectra are performed following the procedure described above. For 1ES 0414$+$009 the intrinsic spectrum is restricted to a power-law due to the limited number of degrees of freedom available. \\
All intrinsic spectral parameters are found to be reasonable, in agreement with typical emission models. Note that this paper is focused on the EBL measurement, and discussion of intrinsic spectral parameters will be detailed in a forthcoming paper.  Individual fit qualities obtained with Eq.~\ref{FitFunction} are shown in Table~\ref{Chi2s}. The fits are not improved when considering more complex intrinsic functions such as ones with cut offs. Low fit qualities can be related to small spectral irregularities that cannot be accounted for in the parameterization. The unfolding covariance matrix can also reduce the fit quality.

\begin {table}[]
\caption{Fit qualities of the different data sets after the joint fit.}
\small
\begin{center}
\begin{tabular}{l @\qquad c}
%\begin{tabular}{l @\qquad S}
%\begin{tabular}{p{3.5cm}p{2cm}p{2cm}p{2cm}p{2cm}}
  \hline
 Data set & $\chi^2_{j, \rm{LP+EBL}}/$ndf \\ 
  \hline
  Mrk 421 (1) & 5.17/5 \\
  Mrk 421 (2) & 17.3/6 \\
  Mrk 421 (3) & 8.83/6 \\
  Mrk 421 (4) & 16.37/5 \\
  Mrk 421 (5) & 6.19/5 \\
  Mrk 501 & 14.18/5 \\
  PKS 2005$-$489 (1) & 13.3/2\\
  PKS 2005$-$489 (2)  & 3.17/5\\
  PKS 2155$-$304 (1) & 9.1/6 \\
  PKS 2155$-$304 (2) &  6.71/4\\
  PKS 2155$-$304 (3) &  11.76/7\\
  PKS 2155$-$304 (4) &  10.3/5\\
  PKS 2155$-$304 (5) & 3.23/7\\
  PKS 2155$-$304 (6) & 4.37/7\\
  PKS 2155$-$304 (7) &  12.29/6\\
  PKS 2155$-$304 (8) & 19.9/7\\
  1ES 0229$+$200 & 2.07/1\\
  H 2356$-$309 &  6.21/3\\
  1ES 1101$-$232 & 1.9/5 \\
  1ES 0347$-$121 &  1.4/3 \\
  1ES 0414$+$009 & 3.2/1 \\
   \hline
\end{tabular}
\end{center}
\label{Chi2s}
\end{table}

The goodness-of-fit estimator is $ \sum_j \chi^2_{j, \rm{LP+EBL}}$. Its value after the joint fit is 176.7.
Considering the four additional EBL degrees of freedom as common parameters and using Wilks' theorem, this can be translated into an EBL detection significance of 9.5$\sigma$ with respect to the log-parabola hypothesis without EBL.

\begin{figure}[]
\centering
\subfigure[\label{EBLRes_SubFig_1}]{\includegraphics[scale=0.29]{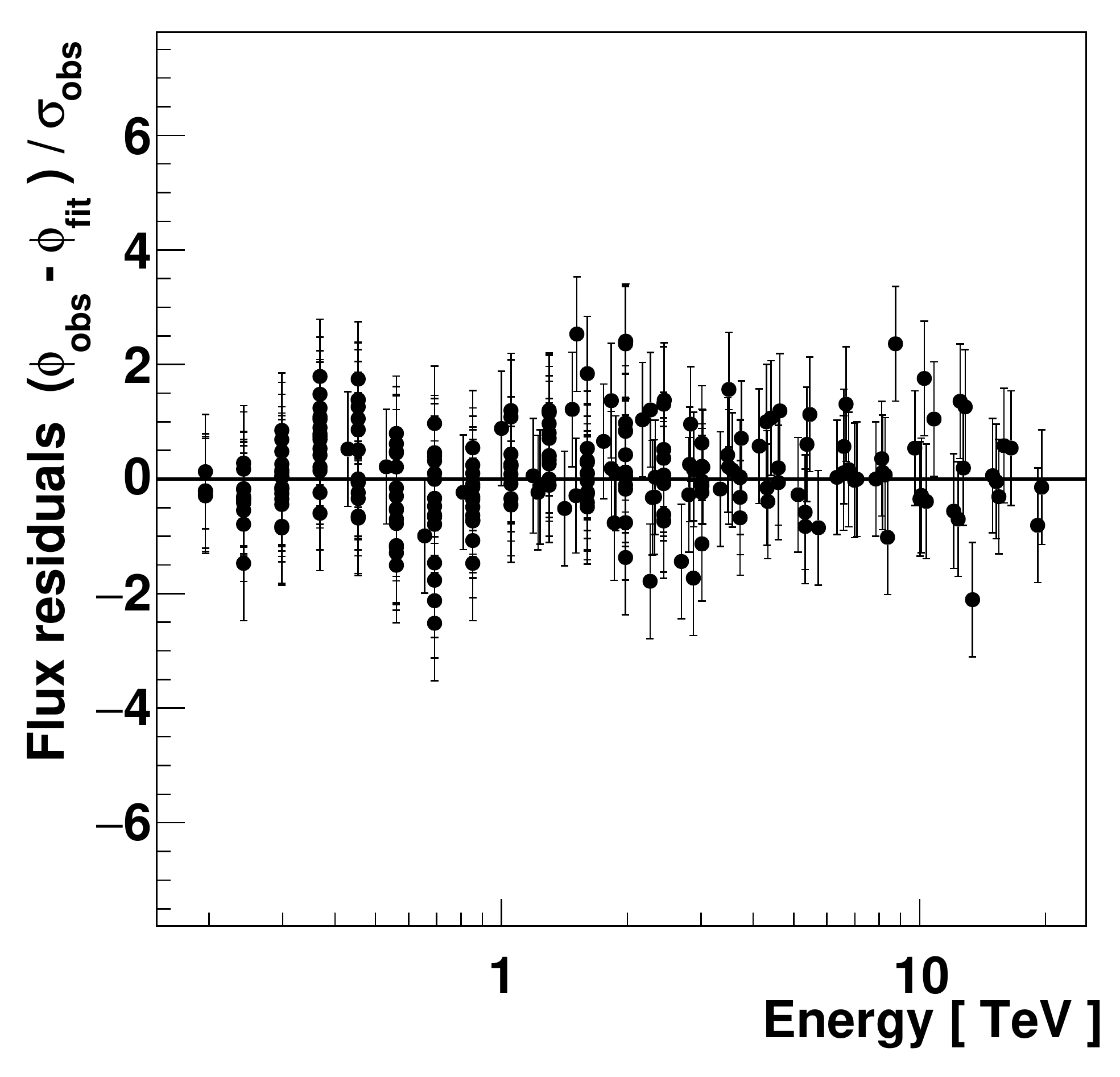}}\quad
\subfigure[\label{EBLRes_SubFig_2}]{\includegraphics[scale=0.29]{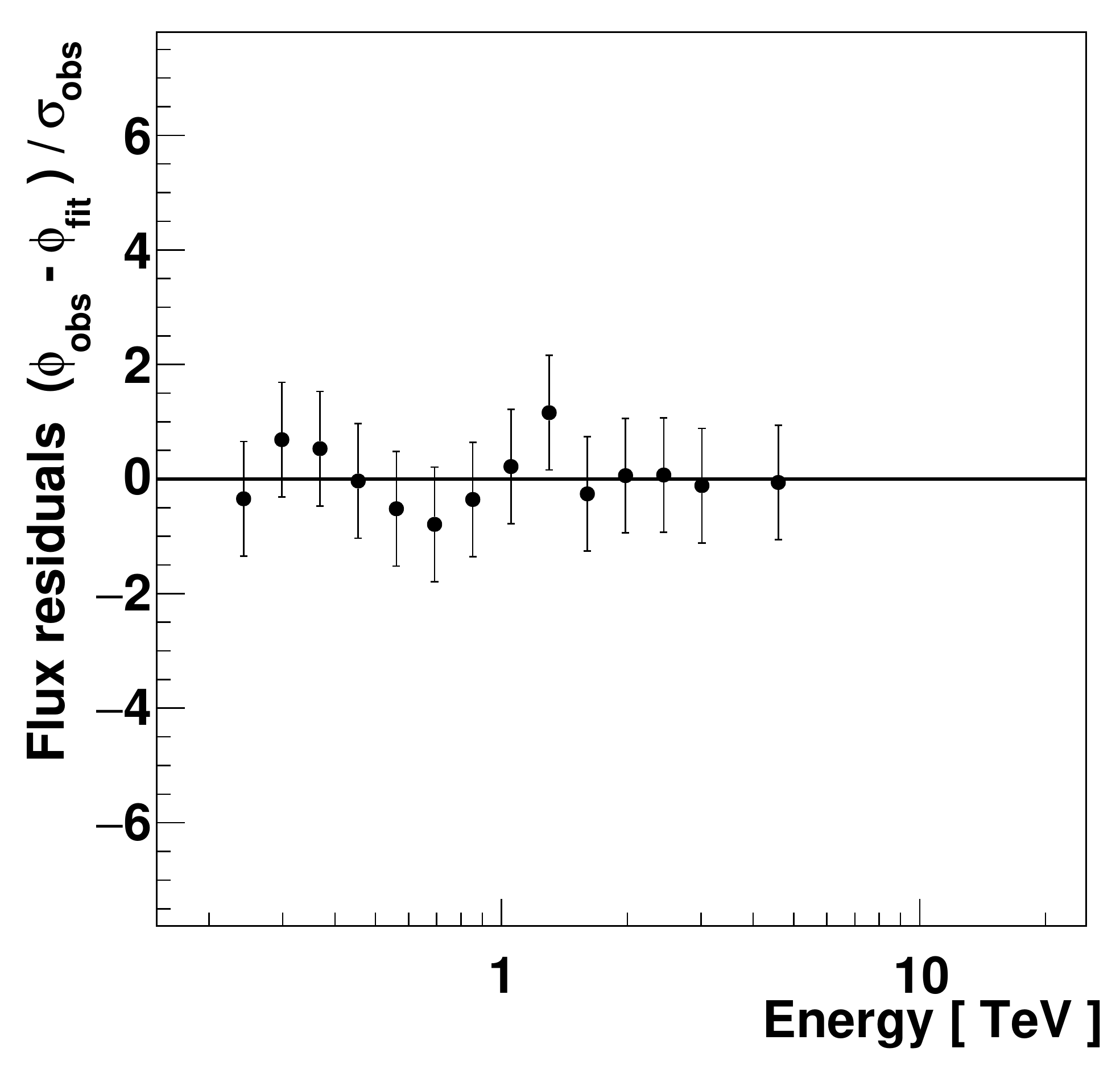}}
\label{EBLResiduals}
\caption{\subref{EBLRes_SubFig_1}~Residuals of the whole sample of spectra for the log-parabola fit with EBL. \subref{EBLRes_SubFig_2}~Example residuals to the log-parabola fit with EBL for the subset PKS~2155$-$304 (5).}
\end{figure}

Figure~\ref{EBLRes_SubFig_1} displays the accumulated residuals from the fit with EBL. The modulation seen in Fig.~\ref{SubFig_2} is reduced, showing that the addition of EBL-related degrees of freedom provides a better description of the data. Figure~\ref{EBLRes_SubFig_2} shows this effect for the subset PKS~2155$-$304~(5), to be compared with Fig.~\ref{SubFig_3}.

\begin{figure*}[]
\center
\includegraphics[scale=0.85]{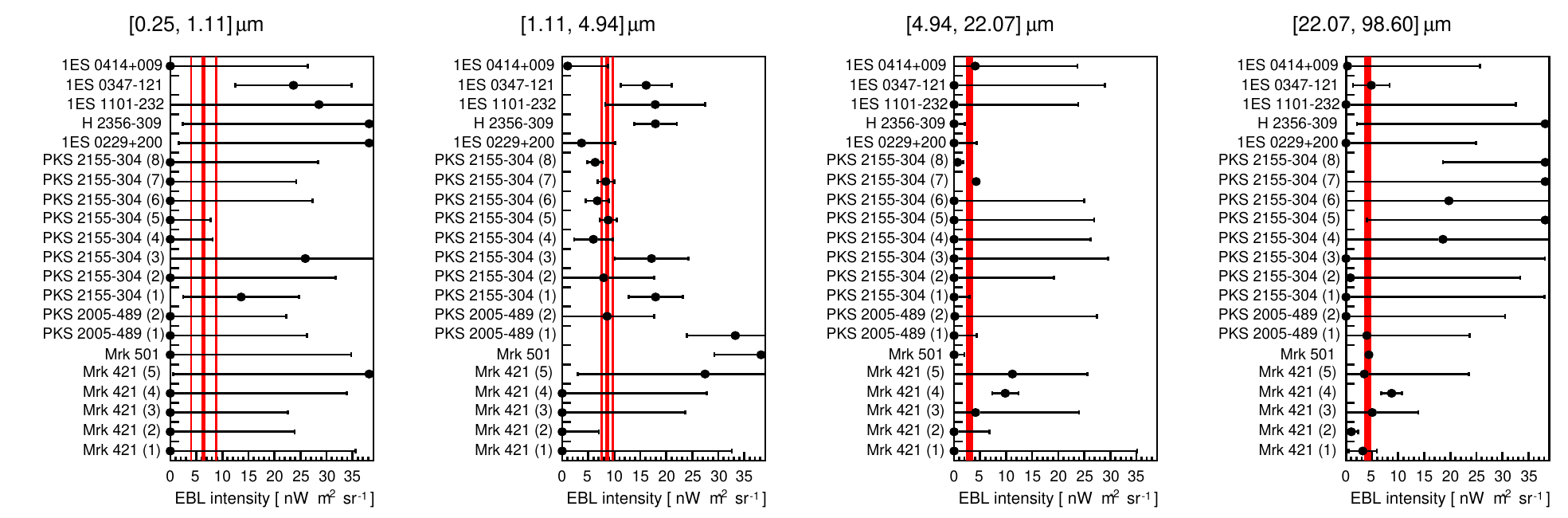}
\caption{Individual EBL levels and errors in the different wavelength bands obtained from the fit of each spectrum. The plain red lines represent the combined values and statistical errors.}
\label{IndividualBinResults}
\end{figure*}

In the following, the results are presented by converting the local EBL energy density $\epsilon_0^2 \frac{dn_{\mbox{\tiny{EBL}}}}{d\epsilon_0}$ (in units of eV m$^{-3}$) into specific intensity $\lambda I_\lambda$ (in units of nW m$^{-2} $sr$^{-1}$) following the relation $ \lambda I_\lambda  = \frac{c}{4 \pi} \epsilon_0^2 \frac{dn_{\mbox{\tiny{EBL}}}}{d\epsilon_0} $. \\

Individual results per data set reflect the relative sensitivity range of the different spectra, as illustrated in Fig.~\ref{IndividualBinResults}. For instance, subsets of PKS 2155$-$304 show optimal sensitivity in the 1.1 -- 4.94 $\mu$m band, with precisely fitted EBL levels. Subsets of Mrk 421 and Mrk 501 lead to precise measurement at larger wavelengths. Alternatively, when the range covered by a given spectrum does not constrain the EBL in a given band, the corresponding uncertainty on the fitted level is large. This behavior shows the ability of the method to probe the different ranges in EBL wavelength depending on the source properties (redshift, accuracy of spectral points measurements, covered energy range). When no clear spectral modulation is identified, a softer or more curved intrinsic spectrum compatible with null levels of EBL can be preferred. This would not be the case if intrinsic curvature was forbidden. The method thus prevents one from interpreting an overall spectral curvature as EBL detection, because its signature -- if present -- is imposed to be a more complex feature.

\subsection{Spectrally resolved EBL levels}
\label{CombineSection}
Individual signals are combined to obtain a collective EBL measurement.
This approach takes advantage of the large sample of high-quality spectra from sources at various redshifts. Indeed, the EBL is expected to have a coherent effect that can be interpreted collectively. 

The combined EBL level in a band is obtained as an error-weighted average over all data sets $\left\lbrace j  \right\rbrace$:
\begin{equation}
\langle \rho_{i} \rangle =\frac{\sum_{j} \rho_{i,j} / \sigma_{\rho_{i,j}}^2}{\sum_j \sigma_{\rho_{i,j}}^{-2}}.
\label{CombineEBL}
\end{equation}
The error on a combined EBL level takes into account both individual errors and the dispersion of the $n$ individual measurements around the averaged value. This is done by weighting the sample variance in each wavelength band by the corresponding reduced $\chi^2$: 
\begin{equation}
\sigma_{\langle \rho_{i} \rangle	}  = \sqrt{\frac{1}{\sum_j \sigma^{-2}_{\rho_{i,j}}}  \frac{1}{(n-1)}  \sum_j \frac{ \left(\rho_{i,j}-\langle \rho_{i} \rangle \right) ^2}{\sigma^2_{\rho_{i,j}}} }.
\label{CombineErrors}
\end{equation}
Following such an approach, the uncertainty on a combined EBL level is slightly corrected to yield conservative results.\\

Individual EBL levels should essentially be compatible with each other as the EBL is assumed to be a diffuse isotropic background. Potential anisotropies  of the EBL \citep{Furniss:2014bna, 2017ApJ...835..237A} are estimated to be beyond the sensitivity of this EBL measurement method. The dispersion of individual values can reflect systematic uncertainties and potential limitations of the procedure. The latter can be due for instance to the fit of patterns in spectra that might not be related to EBL absorption and not accounted for in the intrinsic spectrum parameterization. However, to avoid introducing a bias, once the method is fixed, any kind of  individual tuning of parameters is forbidden, and all fits are performed in one single blind procedure. Equation \ref{CombineErrors} ensures that significant deviations to the average value degrade the precision on the EBL measurement. Of the 84 points displayed in Fig.~\ref{IndividualBinResults}, only a few deviate from the average value. Note that the thin red lines only represent the statistical uncertainties on the combined EBL levels. For the lowest wavelength band (0.25 -- 1.1 $\mu$m), Eq.~\ref{CombineErrors} slightly reduces the size of the statistical uncertainty on the combined level because of the under-dispersion of individual levels due to their very large error bars. The large uncertainty on this combined level due to the poorly constrained individual measurements is properly taken into account with the consideration of systematic uncertainties in Sec.~\ref{Section:Sys}.

The combined EBL levels are summarized in Table~\ref{EBLTable} and shown in Fig.~\ref{EBLStatPlusSys}. Details on the estimation of systematic uncertainties are given in Sec.~\ref{Section:Sys}. Comparisons with various EBL constraints and models are left for the discussion in Sec.~\ref{Discussion}.

{\renewcommand{\arraystretch}{1.2}
\begin {table}[H]
\begin{center}
\small{
%\begin{tabular}{ccc @\quad c @\quad  c @\quad c}
\begin{tabular}{SSS @\quad S @\quad  c @\quad c}
  \hline
 $\lambda$ & $\lambda_{\mathrm{min}}$  & $\lambda_{\mathrm{max}}$  & $\rho$ & $\rho_{\mathrm{min}} \ (\text{sys})$ & $\rho_{\mathrm{max}}\ (\text{sys})$  \\
  \hline
   0.52 & 0.25 & 1.11  & 6.42 & 4.02 (0) & 8.82 (14.5)\\
   2.33 & 1.11 & 4.94 & 8.67 & 7.63 (6.68) & 9.71  (11.80) \\ 
   10.44 & 4.94 & 22.07  & 3.10  & 2.62 (1.16) & 3.59 (4.17)\\ 
    46.6 & 22.07 & 98.60 & 4.17 & 3.71 (2.55) & 4.63 (6.30)  \\ 
   \hline
\end{tabular}
}
\end{center}
\caption{Combined EBL levels ($\rho$, in $ \text{nW} \text{m}^{-2} \text{sr}^{-1}$) in the different wavelength bands ($\lambda$, in $\mu \text{m}$).}
\label{EBLTable}
\end{table}
}

\subsection{Systematic uncertainties}
\label{Section:Sys}
In addition to the previously-mentioned dispersion of individual values, different sources of systematic uncertainties are investigated.\\

Systematic uncertainties related to the EBL evolution hypothesis are estimated considering the toy-model evolution parameterization used in other model-independent approaches to determine the EBL \citep{Meyer:2012us, Biteau:2015xpa}. It consists of a global rescaling of the photon density with respect to the cosmological expansion $(1+z)^{3} \rightarrow (1+z)^{3-f_{\mathrm{evol}}}$, where the value of $f_{\mathrm{evol}}$ is chosen in order to mimic the evolution function of EBL models. Adopting the typical value $f_{\mathrm{evol}}=1.2$ \citep{2008IJMPD..17.1515R} leads to differences less than 5$\%$ in EBL levels with respect to the case where the evolution function is extracted from \citet{Franceschini:2008tp}, lying within statistical errors.

Systematic uncertainties related to the energy scale in \g-ray spectra measurements could originate from variations of the Cherenkov light yield due to, \textit{e.g.}, fluctuations of atmospheric transparency not accounted for in the simulation, mismatches between real and simulated mirror reflectivities, \textit{etc.} \citep{2014_Hahn}. A systematic energy shift of $\pm 15 \%$ is assumed. It represents a conservative estimate of the absolute energy scale uncertainty with \hess \citep{Aharonian:2006pe}. This energy shift is applied at the spectrum level for all data sets and the whole procedure is redone. The EBL wavelength range is shifted accordingly, so that the possibility is left for different EBL levels in different bands to induce identical patterns in spectra. The observed variations on EBL levels are of the order of $10 \%$, symmetric with respect to the central value and global over the wavelength range, with similar goodness of fit in each case.

If the EBL wavelength range is not shifted according to the \g-ray energy scale, a mismatch between the spectrum energy scale and the relative position of the wavelength bands is introduced. This is equivalent to investigating the effects of bin-shifting in the wavelength bands and can lead to significant changes in the measured EBL levels, reflecting the level of degeneracy between intrinsic spectra and fitted EBL levels. The combined systematic effect of shifts in the energy scale, changes in the wavelength range and changes in fitted intrinsic spectra leads to significant variations from 10$\%$ to 70$\%$ of the central EBL level. These uncertainties strongly depend on the band considered and are not symmetric in intensity. In this way, conservative wavelength-dependent systematic uncertainties are obtained.\\

The influence of the width of the wavelength bands and their number is also investigated. In addition to changes that can naturally arise from the variations of the EBL SED over one band, integrated EBL levels can fluctuate due to the different absorption patterns available when using bands of different size and number. The number of bands is limited by the degrees of freedom available in the joint fit of each spectrum. The use of more (and smaller) bands is then only possible for spectra with sufficient degrees of freedom. Using wavelength bands with larger and smaller width the changes in EBL levels are found to be dependent of the wavelength range considered, from negligible variations up to 40$\%$ variations.

An alternative approach fitting simultaneously all spectra with common EBL parameters is also considered. With this global approach the combination of EBL levels obtained from individual spectra is not needed but the information concerning the contribution of each spectrum to the measurement is lost.
Consistent EBL levels are obtained from this global fit. The level in the lowest wavelength band (0.25 -- 1.11 $\mu$m) appears poorly constrained in the global fit and is compatible with a null level of EBL. This lowest wavelength band is at the limit of the sensitivity of this study, as already apparent from the individual measurements shown on Fig.~\ref{IndividualBinResults}.

\begin{figure*}[]
\center
\includegraphics[scale=0.75]{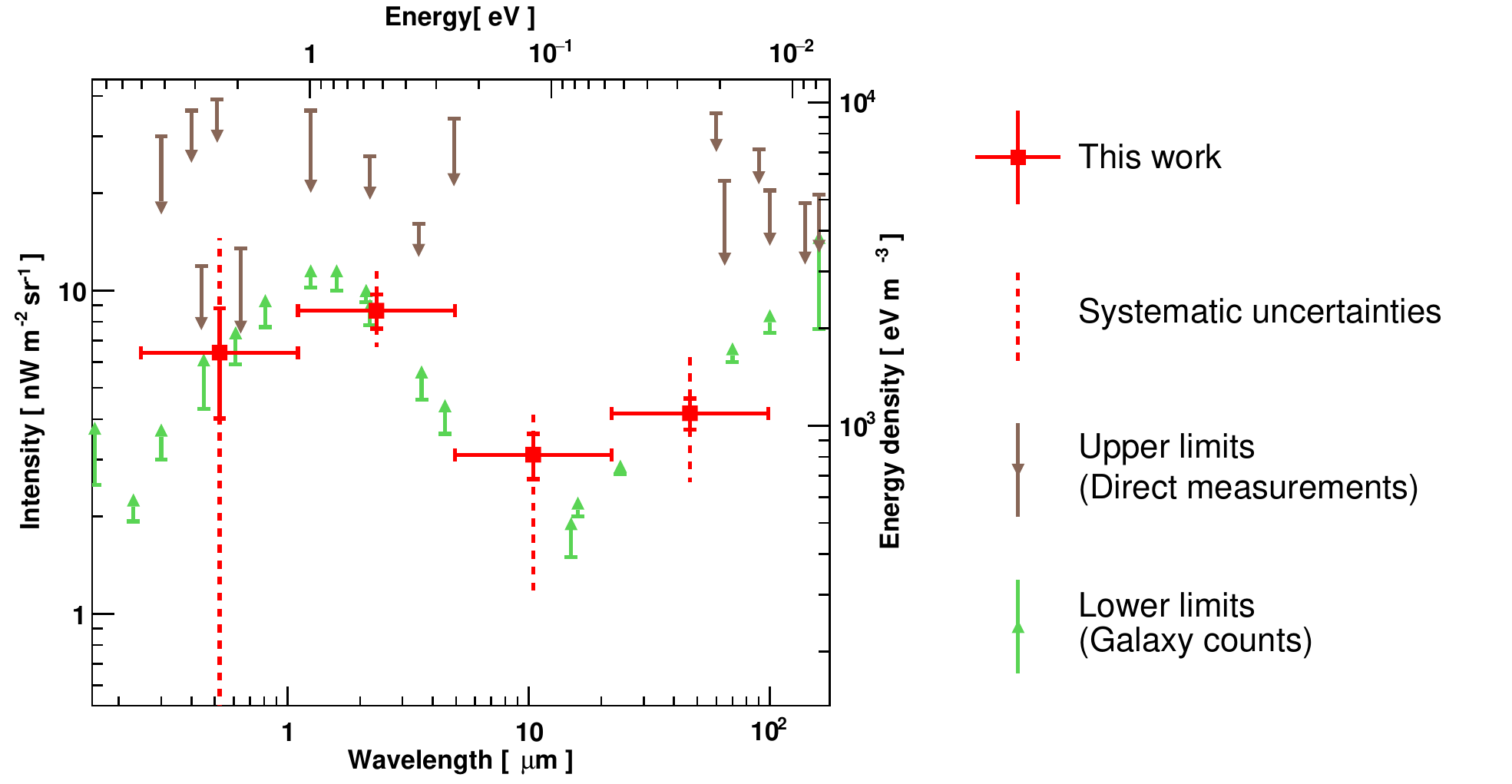}
\caption{Measured EBL spectrum. Obtained levels are represented by the red points. Horizontal lines represent the bandwidth over which the integrated EBL levels apply. Vertical plain lines represent 1-$\sigma$ (statistical) errors from Eq. \ref{CombineErrors}. Dashed lines represent the systematic uncertainties on fitted EBL levels conservatively estimated as explained in the text. Direct constraints on the EBL collected from \citet{Dwek:2012nb} and \citet{Biteau:2015xpa} are shown: lower and upper limits are represented by green and brown arrows respectively.}
\label{EBLStatPlusSys}
\end{figure*}

The envelope of largest variations corresponding to the different kinds of potential systematic errors are represented in Fig. \ref{EBLStatPlusSys} as dashed lines. This behavior shows the relative indeterminacy for the 0.25 -- 1.11 $\mu$m and  4.94 -- 22.07 $\mu$m bands, but also the stronger signals in the 0.25 -- 1.11 $\mu$m and in the 22.07 -- 98.6 $\mu$m bands which are clearly significant beyond systematic uncertainties.

\subsection{Discussion}
\label{Discussion}
\begin{figure*}[]
\center
\includegraphics[scale=0.75]{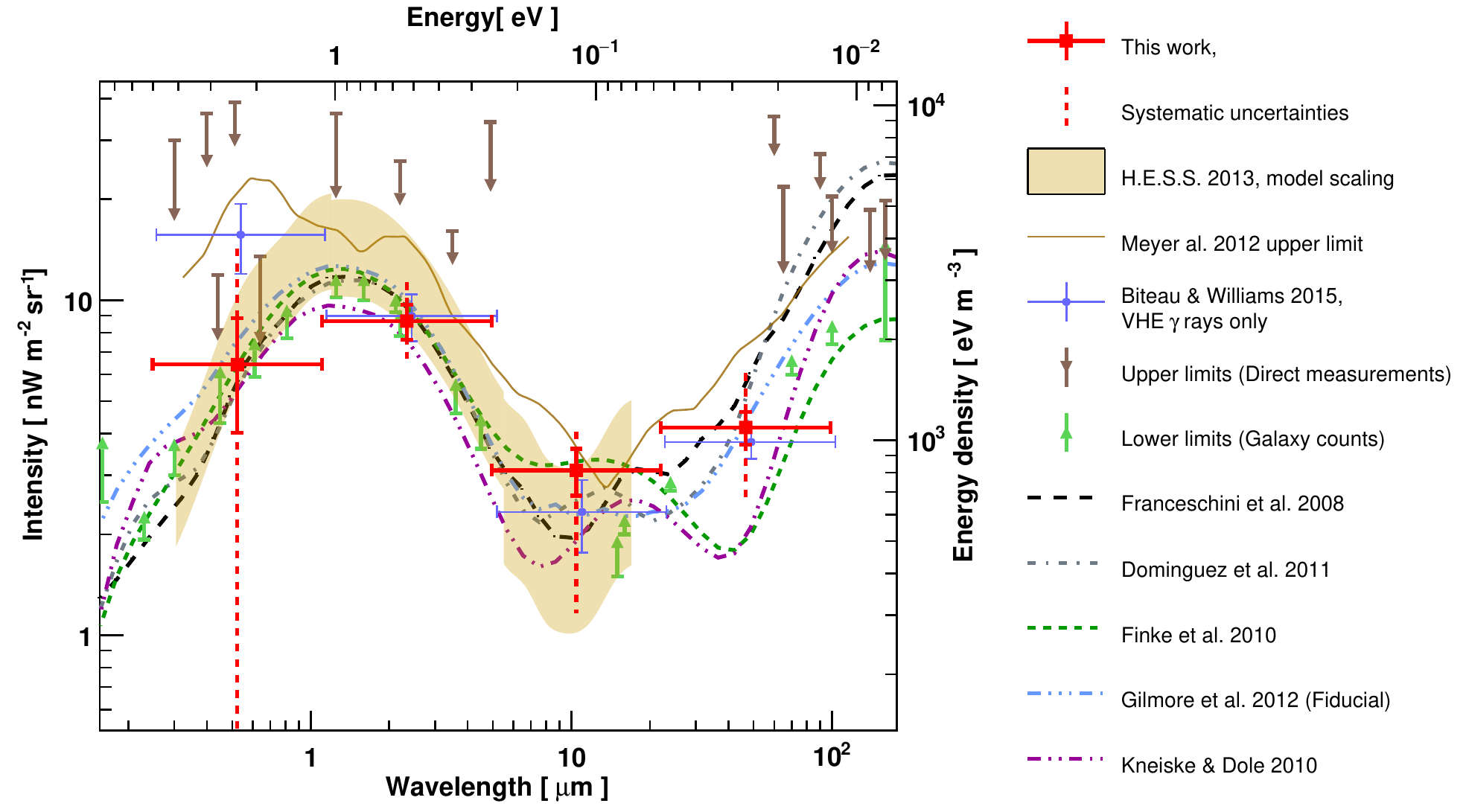}
\caption{Combined EBL levels (red points) compared with various constraints and models. The dashed black line represents the local SED of the model given in \citet{Franceschini:2008tp}, which is used as the template for the HESS2013 model scaling (yellow area). Additional represented models are \citet{Dominguez:2010bv} (grey dotted-dashed line), \citet{Finke:2009xi} (green dashed line), \citet{Gilmore:2011ks} (blue dotted-dashed line), and the lower-limit model of \citet{2010A&A...515A..19K} (purple dotted-dashed line). The model-independent upper limit using VHE and HE data \citet{Meyer:2012us} is shown as a thin brown line. The model-independent measurement of \citet{Biteau:2015xpa} restricted to the use of VHE data is represented by blue points.} 
\label{EBLPlot}
\end{figure*}

Sensitivity to the shape and normalization of the EBL SED is achieved using only VHE spectra obtained with H.E.S.S. Although EBL levels were left free to cover a wide range of possibilities and were thus not constrained in the fits, results are not conflicting with strict lower limits from galaxy counts. This is an interesting point, as the two methods (\g-rays and galaxy counts) are completely independent, and the EBL levels in the present analysis were left free to vary between zero and an arbitrary value.
The obtained results are consistent with state-of-the-art EBL models and are in general agreement with other \g-ray constraints, as shown in Fig.~\ref{EBLPlot}. Again, note that no prior from the displayed models was used in the H.E.S.S. measurement (apart from the evolution factor, which does not strongly influence the $z=0$ results). The obtained results are compatible with the model scaling of HESS2013, although an extended data set is used and that the treatment of the data is very different. The wavelength range probed by HESS2013 was conservatively restricted to the central value of the pair-creation cross section, neglecting its width. In the present study, the wavelength range probed is extended further in the infrared because optical depth values must be computed over the whole kinematically-allowed range for pair-creation with the most energetic \g-rays  of the sample, as described in Sec.~\ref{EBLOpticalDepth}.
The obtained EBL levels close to lower limits in the optical range are in line with \textit{Fermi} LAT results \citep{Ackermann:2012sza} probing the EBL at higher redshifts and at lower wavelengths and also with the upper limit obtained following the detection of the high-redshift quasar PKS~1441+25 at $z=0.94$ by VERITAS \citep{Abeysekara:2015pjl} and MAGIC \citep{Magic1441}.  The results are also in general agreement with other constraints not represented  in Fig.~\ref{EBLPlot} obtained using \g-rays \citep{Abramowski:2013dya, Ahnen:2016gog} or with the results of empirical approaches to the determination of the EBL SED \citep{Helgason:2012xj,Stecker:2016fsg}.

While an important conclusion of this work is to show that \hess \g-ray spectra alone contain enough information to determine the EBL shape and normalization, the sensitivity of such an approach remains limited. The coarse EBL binning achievable and the conservatively estimated uncertainties on EBL levels show that a fine spectroscopy of the EBL SED (resolving fine substructures in the EBL spectrum \textit{e.g.} due to dust sub-components) is out of reach using only present VHE data. The compatibility between H.E.S.S. measurements and the lower limits from galaxy counts does not suggest a transparency anomaly of the universe to VHE \g-rays (as hinted at in \citealt{Horns:2012fx}), for the redshift range considered.

\begin{figure*}[]
\center
\includegraphics[scale=0.75]{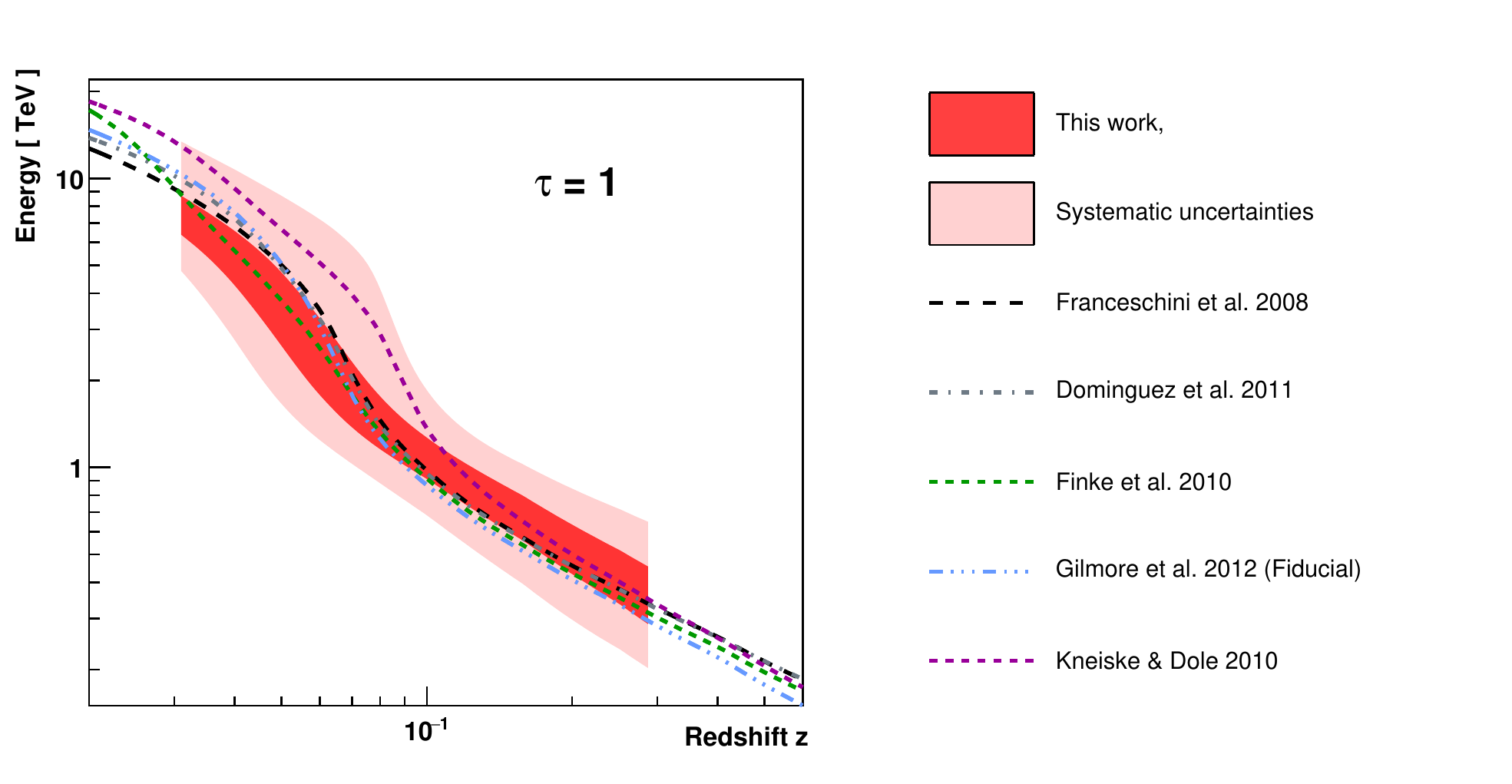}
\caption{Obtained \g-ray horizon for $\tau=1$ in the redshift range covered by the blazar sample and comparison with selected models \citep{Franceschini:2008tp, Dominguez:2010bv, Finke:2009xi, Gilmore:2011ks, 2010A&A...515A..19K}.}
\label{Tau1}
\end{figure*}

The results in terms of EBL intensity can be translated into a corresponding \g-ray horizon. The \g-ray horizon for $\tau = 1$ is a standard illustration of the EBL-absorption effects \citep{Fazio:1970pr}. It corresponds to the typical attenuation length of \g-rays at a given observed energy. The $\tau = 1$ energy-redshift horizon envelopes corresponding to the measured EBL levels and their errors are represented in Fig.~\ref{Tau1}. Iso-$\tau$ curves of selected models are shown for comparison.
These results in terms of \g-ray horizon or optical depths are also compatible with the horizon derived from the SED of state-of-the-art EBL models. Here also the limited sensitivity of the approach appears, as the consideration of systematic uncertainties significantly enlarges the width of the horizon envelope. This shows the difficulty in interpreting such results in terms of transparency anomalies that could be due to second-order propagation effects.

The reduction of uncertainties would of course be possible using additional data and priors. For instance, using data from \textit{Fermi} LAT at lower energy, the degeneracy between intrinsic spectra and EBL absorption can be reduced. This assumes continuity in energy of the intrinsic spectrum behavior, and can also introduce additional systematics due to absolute flux level uncertainties.
Taking into account direct EBL measurements and strict lower limits from galaxy counts can \textit{de facto} restrict the range of variations for EBL levels. Such strong priors were avoided here in order to address the question of EBL information contained only in VHE spectra obtained with H.E.S.S. The detailed study of the intrinsic spectra obtained with this EBL results is left for a dedicated forthcoming paper.

\section{Summary and conclusion}
A determination of the EBL SED with the H.E.S.S. array of Cherenkov telescopes is presented. 
This is achieved using a new method: coherent patterns in the high-quality unfolded spectra of blazars observed by \hess are translated into EBL intensity levels resolved in wavelength, under the assumption that intrinsic spectra are described by smooth concave shapes. The EBL signature is preferred at the 9.5$\sigma$ level compared to the null hypothesis. Combined EBL levels are compatible with current constraints and models, and no indication of an opacity anomaly is found. This robust result demonstrates for the first time the capability of \hess to measure the EBL SED independently of any existing EBL constraints and models.

%\acknowledgments
\begin{acknowledgements}
The support of the Namibian authorities and of the University of Namibia in facilitating the construction and operation of H.E.S.S. is gratefully acknowledged, as is the support by the German Ministry for Education and Research (BMBF), the Max Planck Society, the German Research Foundation (DFG), the Alexander von Humboldt Foundation, the Deutsche Forschungsgemeinschaft, the French Ministry for Research, the CNRS-IN2P3 and the Astroparticle Interdisciplinary Programme of the CNRS, the U.K. Science and Technology Facilities Council (STFC), the IPNP of the Charles University, the Czech Science Foundation, the Polish National Science Centre, the South African Department of Science and Technology and National Research Foundation, the University of Namibia, the National Commission on Research, Science \& Technology of Namibia (NCRST), the Innsbruck University, the Austrian Science Fund (FWF), and the Austrian Federal Ministry for Science, Research and Economy, the University of Adelaide and the Australian Research Council, the Japan Society for the Promotion of Science and by the University of Amsterdam.
We appreciate the excellent work of the technical support staff in Berlin, Durham, Hamburg, Heidelberg, Palaiseau, Paris, Saclay, and in Namibia in the construction and operation of the equipment. This work benefited from services provided by the H.E.S.S. Virtual Organisation, supported by the national resource providers of the EGI Federation.
\end{acknowledgements}

\bibliography{eblpaperbib}
\bibliographystyle{aa}
%\clearpage
\end{document}